\documentstyle[twoside,fleqn,espcrc2]{article}

\newcommand{\AmS}{{\protect\the\textfont2
  A\kern-.1667em\lower.5ex\hbox{M}\kern-.125emS}}

\def\cmdue {cm$^{-2}$}
\def\ltsima{$\; \buildrel < \over $\sim$ \;$}
\def\lsim{\lower.5ex\hbox{\ltsima}}
\def\gtsima{$\; \buildrel > \over $\sim$ \;$}
\def\gsim{\lower.5ex\hbox{\gtsima}}
\def\spose#1{\hbox to 0pt{#1\hss}}
\def\approxlt{\mathrel{\spose{\lower 3pt\hbox{$\sim$}}
        \raise 2.0pt\hbox{$<$}}}
\def\approxgt{\mathrel{\spose{\lower 3pt\hbox{$\sim$}}
        \raise 2.0pt\hbox{$>$}}}

\def\aa #1 #2 {A\&A, {#1}, #2}
\def\mon #1 #2 {MNRAS, {#1}, #2}
\def\apj #1 #2 {ApJ, {#1}, #2}
\def\nat #1 #2 {Nature, {#1}, #2}
\def\pasj #1 #2 {PASJ, {#1}, #2}

\title{Discovery of X--rays from the  supernova remnant G0.9+0.1}

\author{L. Sidoli\address{Dipartimento di Fisica, Universit\`a di Milano, 
	Via Celoria 16, I-20133 Milano, Italy}, %
	G.L. Israel\address{Osservatorio Astronomico di Roma, 
	Via dell'Osservatorio 2,
        I-00040 Monteporzio Catone (Roma), Italy}, %
        L. Chiappetti\address{Istituto di Fisica Cosmica del C.N.R., 
         Via Bassini 15, I-20133 Milano, Italy}, %
        A. Treves$^{a}$, %    
        M. Orlandini\address{TeSRE, 
	Via P. Gobetti 101, I-40129 Bologna, Italy}, %
        E. Kuulkers\address{University of Oxford, Nuclear and Astrophysics Laboratory, Keble Road, Oxford OX1 3RH, UK}, % 
        P. Predehl\address{MPE, Garching, Germany}, %
        J. Heise\address{SRON-Utrecht,      
	Sorbonnelaan 2, 3584 CA Utrecht, The Netherlands}
        and
	S. Mereghetti$^{c}$%
}
\input psfig.tex       
\begin{document}

\begin{abstract}
During the $BeppoSAX$ survey of the Galactic Center region, we have discovered  X--ray emission from the central region of the supernova remnant G0.9+0.1.
The high interstellar absorption ($N_H\sim3\times10^{23}$\cmdue) is consistent with a 
distance of order of 10 kpc and, correspondingly, an X--ray luminosity of 
 $\sim$10$^{35}$ erg s$^{-1}$.

Although we cannot completely rule out a thermal origin of the X--ray emission, its small angular extent (radius $\sim2'$), the good fit with a power law, the presence of a flat spectrum radio core, and the estimated SNR age of 
a few thousand years, favour the interpretation in terms of synchrotron emission
powered by a young, energetic pulsar.
 
\end{abstract}

\maketitle

\section{THE BeppoSAX GALACTIC CENTER SURVEY}
 
The BeppoSAX satellite \cite{BA97} is performing a survey of the Galactic Center region using its Narrow Field Instruments. 

The MECS instrument \cite{BB97}, with its good imaging capabilities up to energies of $\sim$ 10 keV, is particularly indicated to study this crowded and highly absorbed region of the sky.

Figure 1 shows a mosaic of the MECS images obtained during the first year of this project.
Several known point sources are visible (including one at the position of SgrA*),
as well as diffuse emission.
Here we concentrate on the discovery of X--ray emission from G0.9+0.1, one of the radio supernova remnants in the Galactic Center region.

\begin{figure*}
\centerline{\psfig{figure=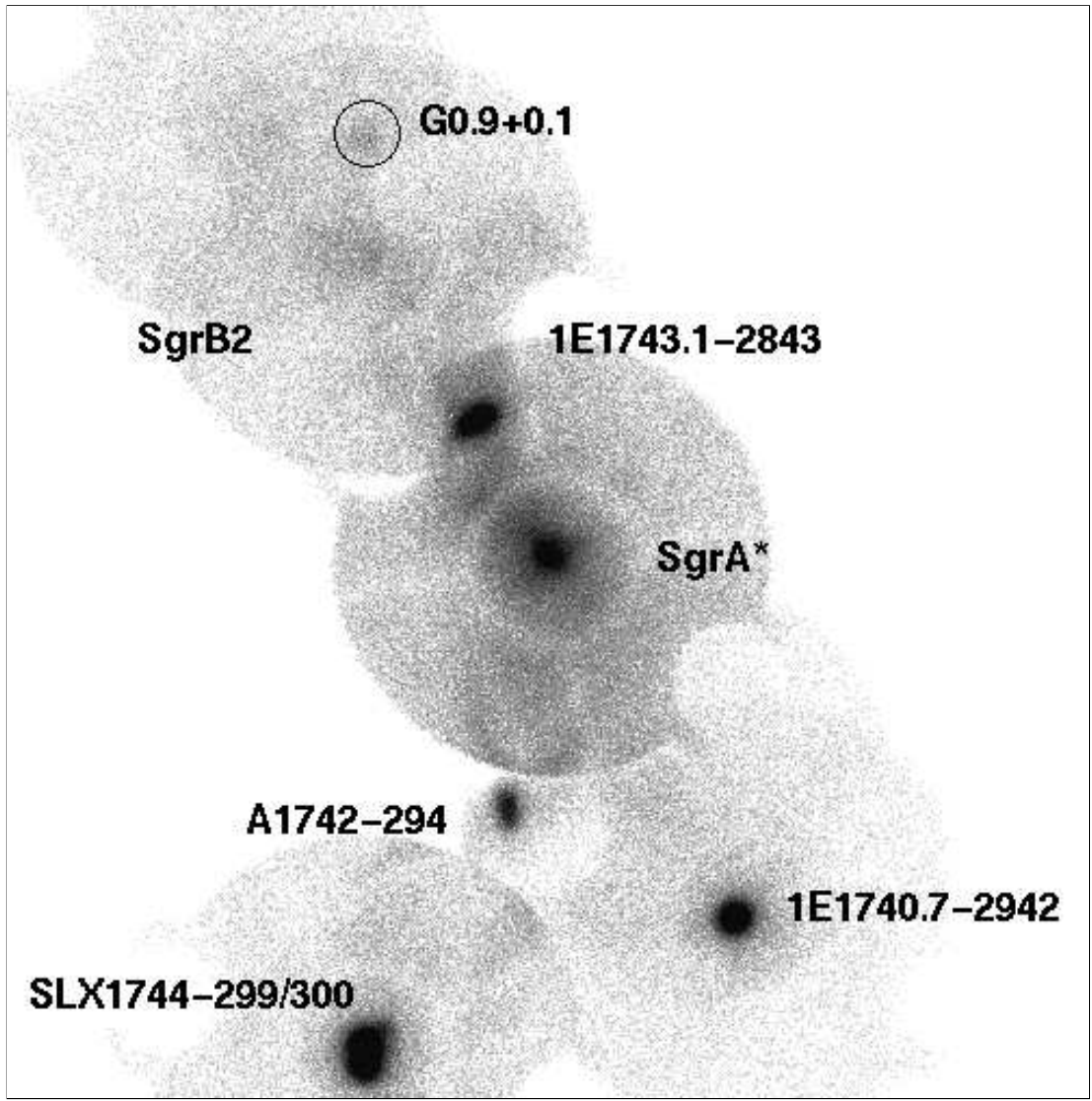,width=15cm,height=15cm} }
\caption{$BeppoSAX$ image  of the Galactic Center Region   ($-1.5<l<1.5$) in the 2-10 keV energy range.}
\label{fig:largeenough}
\end{figure*}
 
%
%\begin{figure}[htb]
%\framebox[55mm]{\rule[-21mm]{0mm}{43mm}}
%\caption{Remember to keep details clear and large enough.}
%\label{fig:toosmall}
%\end{figure}

\section{OBSERVATIONS AND RESULTS}

A new source, located about 14 arcmin north of the molecular cloud SgrB2 
and coincident with the radio supernova remnant G0.9+0.1 was discovered 
during a 49 ksec long pointing performed in April 1997 \cite{M97}. The source 
was reobserved for 51 ksec in September 1997, but only with two MECS, due 
to the failure of MECS1 in May 1997.

The source flux was consistent with a constant value in both observations.
The spectra extracted by combining both observations are equally well 
fitted by power law, blackbody and thermal bremsstrahlung models, 
while a Raymond-Smith thermal plasma model 
with abundances fixed at the solar values gives a slightly worse result ($\chi^2=1.20$). The best fit parameters, similar to those derived from the April observation alone \cite{M97}, are given in Table 1.

All the models give values of $N_{H}$ greater than $2~10^{23}~$\cmdue, 
indicating a distance of several kiloparsecs, and thus supporting the identification with G0.9+0.1 that is probably close to the Galactic Center
or even beyond it \cite{D71}.

Assuming a distance of 10 kpc, the luminosity in the 2--10 keV range is
$L_{X}=1.5\times10^{35} d_{10}^{2} $ erg  s$^{-1}$, for the power law best fit.

A search for pulsations for periods in the range 2048 s to 8 ms gave a negative result,
with the following
upper limits on the pulsed fraction: 53\% for 8 $<$ P $<$ 16 ms, 
38\% for 16 $<$ P $<$ 32 ms, and 33\% for P $>$ 32 ms.

\section{G0.9+0.1: A COMPOSITE RADIO/X--RAY SUPERNOVA REMNANT}
 
In the radio band G0.9+0.1 consists of a steep-spectrum radio shell 
of $\sim$ $8'$ diameter surrounding a core component with a flatter spectrum and significant polarization \cite{HB87,G94}.

Thus G0.9+0.1 belongs to the class of ``composite'' supernova remnants \cite{W88}, that,
in addition to the radio/X--ray shell formed by the expanding ejecta, 
show the signature of a central neutron star powering non--thermal emission through the loss of rotational energy.

Due to the high absorption along this direction, G0.9+0.1 was not detected 
with ROSAT.   Our $BeppoSAX$ observations provide the first information on the X--ray
emission from G0.9+0.1 (only a questionable marginal detection with the
Einstein satellite IPC \cite{HB87} had been previously reported).

The peak of the X--ray emission is coincident with the SNR radio core and
there is no evidence for a spatial extension greater than the instrumental resolution at this off-axis angle (R $<$ $2'$).
Therefore, we are clearly seeing X--rays emitted from the central region 
of the remnant and not from the $8'$ shell
that would appear clearly resolved in the MECS images.
  
Some SNRs, like for example W44 \cite{RR}, present a centrally peaked X--ray emission 
of thermal origin. The thermal nature of the emission
is clearly demonstrated by the detection of lines in
their X--ray spectra. 
All the SNRs of this kind have a limb-brightened radio
morphology  without  a flat--spectrum core, contrary to the case of G0.9+0.1. 
Also considering that  the thermal plasma model
gave the worst fit to our  data,   
we favour the alternative interpretations
related to the likely presence of a neutron star
at the center of G0.9+0.1.

\subsection{Thermal Emission from a Neutron Star?}

The blackbody spectral fit imply an emitting surface with radius $R=0.3^{+0.3}_{-0.1} d_{10}$ km, 
consistent with emission from a small polar cap region, hotter than 
the rest 
of the neutron star due to anisotropic heat diffusion from the interior 
and/or to reheating by relativistic particles backward accelerated 
in the magnetosphere \cite{HR93}.

In general, this should produce a periodic flux modulation, but the strong
gravitational bending effects severely reduce the observed pulsed fractions \cite{P95}. Our upper limits on the possible flux modulations are not strong enough 
to pose serious problems to this interpretation. 
However, the fitted blackbody temperature is higher than that observed in all
the other X--ray emitting radio pulsars \cite{Be}.

\begin{table*}[hbt]
% space before first and after last column: 1.5pc
% space between columns: 3.0pc (twice the above)
\setlength{\tabcolsep}{1.5pc}
% -----------------------------------------------------
% adapted from TeX book, p. 241
\newlength{\digitwidth} \settowidth{\digitwidth}{\rm 0}
\catcode`?=\active \def?{\kern\digitwidth}
% -----------------------------------------------------
\caption{{\bf} Results of the Spectral Fits (errors are 90\% c.l.).}
\label{tab:models}
\begin{tabular*}{\textwidth}{@{}l@{\extracolsep{\fill}}ccccc}
\hline
Model &	Column density         & Parameter & Red. $\chi^2$  &Flux  (2--10 keV)   \\
      & ($10^{23}$ cm$^{-2}$)  &           &        (49 d.o.f.)       & ($10^{-11}$ ergs cm$^{-2}$ s$^{-1}$)   \\
\hline
Power law        & $3.4^{+1.6}_{-0.9}$    & $\alpha=3.7^{+1.3}_{-1.0}$              & $0.70$    & $1.36^{+3.74}_{-0.72}$ \\
Bremsstrahlung   & $2.8^{+0.2}_{-0.8}$      & $T_{\rm br}=3.2^{+2.3}_{-0.7}$ keV    & $0.71$    & $0.71^{+0.14}_{-0.26}$ \\
Black body       & $2.2^{+0.8}_{-1.2}$      & $T_{\rm bb}=1.2^{+0.3}_{-0.1}$ keV    & $0.74$    & $0.41^{+0.14}_{-0.16}$ \\
Raymond--Smith   & $4.4^{+1.6}_{-0.9}$      & $T_{\rm RS}=1.1^{+0.2}_{-0.2}$ keV    & $1.20$    & $4.2^{+6.8}_{-2.3}$ \\
\hline
%\multicolumn{5}{@{}p{120mm}}{Reprinted from: G.M. Ritcey,
%                             Tailings Management,
%                             Elsevier, Amsterdam, 1989, p. 635.}

\end{tabular*}
\end{table*}

\subsection{An X--Ray Synchrotron Nebula?}

A different explanation involves non-thermal emission powered by the 
rotational energy loss of a relatively young neutron star.
The radio shell radius of $\sim$10 pc implies a lower limit to the 
remnant age of
$\sim$1100 yr, for a free-expansion phase with 
v$\sim$10$^{4}$ km s$^{-1}$. 
If the remnant is expanding adiabatically, from the Sedov model 
we have a shell radius $R\sim14(E_{51}/n_{o})^{1/5}t_{4}^{2/5}$pc.
For typical values $(E_{51}/n_{o})^{1/5}\sim 1$, we derive an age 
of 6,800 years.

Our best fit power law photon index 3.7 is rather steep, compared
to other X--ray synchrotron nebulae. A more typical value of 2
is also consistent with our data within 99\% confidence level (for $N_H\sim2\times10^{23}$\cmdue).
The corresponding X--ray luminosity (2--10 keV), 
 $L_{X}=4.4\times10^{34} d_{10}^{2} $ erg  s$^{-1}$, is 
within the range observed in the central components of other SNRs and 
can be easily powered by a neutron star with an age of a few thousand years.

\section{CONCLUSIONS}

The $BeppoSAX$ discovery of X--ray emission from the central region of G0.9+0.1 has confirmed its plerionic morphology derived
from the radio observations.
The high interstellar absorption is consistent with a distance 
of the order of 10 kpc and, correspondingly, an X--ray luminosity of 
$\geq 10^{35} d_{10}^{2} $ erg  s$^{-1}$.

Although we cannot completely rule out a thermal origin of the X--ray emission,
its small angular extent, the good fit with a power--law, the presence of a flat spectrum radio core, and the estimated SNR age of a few thousand years,
favour the interpretation in terms of synchrotron emission.

High angular resolution observations with AXAF and XMM can test this interpretation
and possibly lead to the discovery of a relatively young, energetic pulsar in G0.9+0.1.

\end{document}